# Exploring shared genetic bases and causal relationships of schizophrenia and bipolar disorder with 28 cardiovascular and metabolic traits

Hon-Cheong So[1,2*], Carlos Kwan-Long Chau[1], Fu-Kiu Ao[1], Cheuk-Hei Mo[7], Pak-Chung Sham[3,4,5,6]

[1]School of Biomedical Sciences, The Chinese University of Hong Kong, Shatin, Hong Kong
[2]KIZ-CUHK Joint Laboratory of Bioresources and Molecular Research of Common Diseases, Kunming Institute of Zoology and The Chinese University of Hong Kong
[3]Department of Psychiatry, University of Hong Kong, Pokfulam, Hong Kong
[4]Centre for Genomic Sciences, University of Hong Kong, Pokfulam, Hong Kong
[5]State Key Laboratory for Cognitive and Brain Sciences, University of Hong Kong, Pokfulam, Hong Kong
[6]Centre for Reproduction, Development and Growth, University of Hong Kong, Pokfulam, Hong Kong
[7]Faculty of Medicine, Chinese University of Hong Kong

**Correspondence to: Hon-Cheong So**, MBBS, PhD, Lo Kwee-Seong Integrated Biomedical Sciences Building, The Chinese University of Hong Kong, Shatin, Hong Kong. Tel: +852 3943 9255; E-mail: hcso@cuhk.edu.hk

**Abstract**

Background: Cardiovascular diseases (CVD) represent a major health issue in patients with schizophrenia (SCZ) and bipolar disorder (BD), but the exact nature of cardiometabolic (CM) abnormalities involved and the underlying mechanisms remain unclear. Psychiatric medications are known risk factors, but it is unclear whether there is a connection between the psychiatric disorders (SCZ/BD) themselves and CM abnormalities.

Objective: To examine polygenic associations between SCZ/BD with CM traits, explore possible causal relationships between them, and identify shared genetic loci and pathways.

Design: Using polygenic risk scores (PRS) and LD score regression, we investigated the shared genetic bases of SCZ and BD with a panel of 28 CM traits. We performed Mendelian randomization (MR) to elucidate casual relationships between the two groups of disorders. We also identified the potential shared genetic variants by a statistical approach based on local true discovery rates, and inferred the pathways involved.

Setting: The analysis was based on large-scale meta-analyses of genome-wide association studies (GWAS).

Participants: GWAS data for SCZ and BD was taken from large meta-analyses (N=82315 and 40225 respectively). We also included GWAS data from 28 CM traits, with total number of subjects exceeding a million.

Main Outcomes and Measures: We found polygenic associations of SCZ with glucose metabolism abnormalities, adverse adipokine profiles, increased wait-hip ratio and raised visceral adiposity. However, BMI showed inverse genetic correlation and polygenic link with SCZ. On the other hand, we observed polygenic associations with an overall favorable CM profile in BD. MR analysis showed that SCZ may be causally linked to raised triglyceride and that lower fasting glucose may be linked to BD, although MR did not reveal other significant causal relationships in general. We also identified numerous SNPs and pathways shared between SCZ/BD with CM traits, some of which are related to inflammation or the immune system.

Conclusions and Relevance:
In conclusion, SCZ patients may be genetically associated with several CM abnormalities independent of medication side-effects, and proper surveillance and management of CV risk factors may be required from the onset of the disease. On the other hand, CM abnormalities in BD are more likely to be secondary.

**Introduction**

Increased rates of cardiovascular diseases (CVD) have become a major area of concern for patients with schizophrenia (SCZ) or bipolar disorder (BD) [1,2]. People with SCZ have a life expectancy around 15 to 20 years shorter than the average population while the life expectancy for bipolar patients is 10 to 15 years shorter [3]. The mortality ratio in SCZ and bipolar patients is about 2.5 to 3-fold as compared to the general population [4]. Deaths from cardiovascular diseases have been proposed as a major contributor to the increased mortality [1,3]. A 3-fold increase in CVD mortality was reported in patients with severe mental illness (schizophrenia-spectrum or bipolar disorders) aged 18-49 and a 2-fold increase in patients aged 50-75 [5].

The metabolic syndrome is a key risk factor for cardiovascular morbidity and mortality. It represents a cluster of metabolic abnormalities including dyslipidemia, impaired glucose tolerance, insulin resistance, hypertension and central obesity [6].

A raised prevalence of metabolic syndrome has been observed in SCZ patients. In the CATIE study [7], the baseline prevalence of metabolic syndrome was 35% and 50% in males and female patients respectively. Mitchell et al. [8] conducted a large-scale meta-analysis comprising 25962 subjects and reported that metabolic syndrome is found in almost 1 out of 3 unselected SCZ patients. For BD, heightened risks of metabolic abnormalities have also been recognized. A recent meta-analysis [9] estimated that the overall rate of metabolic syndrome is 37.3%, which is significantly higher than the general population.

The underlying causes for increased risk of metabolic disorders are not completely understood. It has been suggested that a variety of factors including tobacco smoking, physical inactivity, inadequate somatic health-care services, antipsychotic medications and underlying genetics may all contribute to the heightened metabolic risks [1]. In particular, anti-psychotic medications and mood stabilizers have been established and are widely known as significant contributors to metabolic abnormalities [10,11]. Nevertheless, metabolic abnormalities have also been observed in SCZ patients who are drug-naïve. For example, drug-naïve patients were reported to show impaired glucose tolerance, raised fasting glucose levels and insulin resistance in several studies [12]. Impaired glucose tolerance has also been observed in first-degree relatives of affected patients [13]. Recently a few meta-analyses [14-16] have studied glucose abnormalities among first episode drug-naïve SCZ patients. All of them concluded that SCZ patients had significantly worse glucose profiles than controls. As for other metabolic abnormalities that have been studied, a recent meta-analysis reported elevated triglycerides in first-episode psychosis patients but lower low-density cholesterol (LDL) and total cholesterol (TC) in patients [17]. A thorough review of metabolic syndrome in drug-naïve and treated SCZ patients can be found in Chadda et al [12].

Very few studies have investigated metabolic traits in drug-naïve bipolar patients and compared them to a control population. A recent study showed an increased rate of insulin resistance in these patients but there were no difference in the rate of metabolic syndrome [18]. We are not aware of any systematic review or meta-analysis of cardiometabolic traits in bipolar patients.

Previous individual studies are usually of small sample size (mostly less than 100 subjects), and often included only a subset of metabolic parameters. Although a few meta-analyses have been performed, the range of traits studied are limited (mainly glucose and lipid traits) and heterogeneity among different studies were inevitable. For example, there are often variability in inclusion and exclusion criteria. The duration of untreated psychosis were variable and not recorded in a number of studies. In some studies, patients may have received other psychotropic agents (e.g. mood stabilizer and antidepressants) that may affect metabolic profiles[11,19]. In other studies patients may have been prescribed antipsychotics for a short period of time (usually < 2 weeks)[14]. As a result, confounders such as psychiatric medications may still affect the results and it is hence difficult to confirm whether SCZ or BD itself is truly associated with CM abnormalities. As (germline) genetic variations are not affected by drugs, exploration of shared genetic bases may be more useful in discerning whether the disorders (SCZ/BD) themselves contribute to CM abnormalities in patients. Also, the current study included results large samples of patients, totaling over a million participants. As another limitation of prior studies, *causal* relationships between SCZ/BD and cardiometabolic traits are unclear, due to cross-sectional nature of many studies and other confounding factors.

In view of limitations of previous works, further studies are warranted to investigate potential shared underlying pathophysiology of cardiometabolic traits with SCZ and BD. As both psychiatric and cardiometabolic traits are highly heritable [20,21], it is reasonable to suspect a shared genetic basis between them. In this study we investigated potential shared genetic predispositions by polygenic risk scores (PRS) and a two-dimensional version of the local true discovery rate (tdr) approach[22]. We also performed Mendelian randomization (MR) analysis to delineate whether there are any causal links among the two types of disorders.

In a related work, Andreassen et al.[23] proposed a conditional false discovery rate approach to identify loci associated with both SCZ and CVD risk factors, with an aim to improving gene discovery in SCZ. In the current work we took a different and more comprehensive approach. Compared to the study by Andreassen et al., we employed methodologies based on PRS and LD score regression to assess overall genetic sharing. We also studied the genetic sharing of cardiometabolic risks with BD, a topic not investigated before. In addition, a much wider range of metabolic parameters and diseases were included here, and the sample size of SCZ GWAS in this study reaches ~82000, about 4 times the size of the previous one[24]. Importantly, we also looked into the *directions* of associations in this study which was not addressed previously [23].

The overall analytic workflow is briefly described as follows (Figure 1). Firstly, we made use of large-scale GWAS meta-analyses results for SCZ, BD and a comprehensive panel of 28 metabolic and cardiovascular traits to test for associations of polygenic scores. Linkage disequilibrium (LD) score regression was conducted as well. We also examined PRS associations with cardiovascular risk factors in the Northern Finland Birth Cohort with individual genotype and phenotype data. We then performed MR analysis to assess causal relationship between the two groups of disorders. Finally, we "zoomed in" to discover the genetic variants shared between SCZ and BD with each metabolic trait, and inferred the likely involved pathways by gene-set analyses.

To our knowledge, this is the first systemic and the most comprehensive study to date on the shared genetic bases of cardiometabolic traits with both SCZ and BD using PRS, and the first to uncover shared genetic variants between BD and cardiometabolic diseases with large-scale GWAS data. Except for a recent study which employed MR to study the relationship between BMI and psychiatric disorders[25], we are unaware of other works studying causal links of SCZ/BD with CM traits.

**Materials and methods**

*GWAS samples*

We derived PRS from summary statistics of GWAS meta-analyses. The summary results for SCZ [26] was based on a large-scale meta-analysis ($N$ = 82315) obtained from the Psychiatric Genomics Consortium website (https://www.med.unc.edu/pgc). Summary statistics for BD was based on the study by Hou et al.[27] ($N$ = 40255) available from https://www.ebi.ac.uk/gwas/downloads/summary-statistics. We also obtained GWAS summary statistics for a range of CM traits, including low-density lipoprotein (LDL) ($N$ = 89872), high-density lipoprotein (HDL) ($N$ = 94294), total cholesterol (TC) ($N$ = 94577), triglycerides (TG) ($N$ = 90996) [28], body mass index (BMI) ($N$ = 339224) [29], fasting glucose (FG) ($N$ = 58074)[30], fasting insulin (INS) ($N$ = 51570)[30], waist-hip ratio (WHR) ($N$ = 144446)[31], type 2 diabetes (two datasets, $N$ = 110452 and 16580 respectively)[32,33], coronary artery disease (CAD) ($N$ = 184305)[34], leptin ($N$ = 32161)[35], adiponectin ($N$ = 45891)[36], systolic and diastolic blood pressure (SBP/DBP) ($N$ = 146562)[37], body fat percentage ($N$ = 100716)[38], subcutaneous adipose tissue volume (SAT) ($N$ = 18247), visceral adipose tissue volume (VAT) ($N$ = 18332), pericardial fat volume (PAT) ($N$ = 12204), subcutaneous adipose tissue attenuation (SATHU) ($N$ = 12439), visceral adipose tissue attenuation (VATHU) ($N$ = 12519) and VAT/SAT ratio ($N$ = 18191)[39]. For FG, INS, WHR, leptin, VAT, VAT/SAT ratio, we obtained the summary statistics both with and without adjustment for BMI. For PAT, we included test statistics adjusted for height and weight. Details of individual studies are given in the above references. The summary statistics for blood pressure was taken from exome-based studies[37] which provided the directions of associations. The other publicly available GWAS dataset on BP[40] did not specify association directions. Summary statistics were downloaded via links in LD-hub (http://ldsc.broadinstitute.org/) or the GRASP webpage (https://grasp.nhlbi.nih.gov/FullResults.aspx).

*Derivation of polygenic risk scores (PRS) with summary statistics*

PRS can be formulated as a weighted sum of allelic counts, with the weights determined by the effect sizes of individual genetic variants. The PRS for an individual $i$ ($r_i$) can be denoted as

$$r_i = \sum_i w_j x_{ij}$$

where $x_{ij}$ is the (centered) allelic count for the $j$th SNP for the $i$th individual, $w_j$ is the weight given to the

*j*th SNP, given by the log odds ratio or regression coefficient in a regression. We employed two approaches for PRS testing. In the first approach, only the summary statistics of each pair of traits are used. Association testing was carried out by the method "gtx" first described by Johnson [41]. Details of this methodology was also described in several other studies[42-44]. The method tested for association of the PRS derived from the first trait ($r_i$) with an outcome $y_i$ (with values centered) by regression: $y_i = \alpha r_i$, where $\alpha$ is the regression coefficient. $\alpha$ can be estimated by

$$\tilde{\alpha} = \frac{\sum_i w_j \hat{\beta}_j s_j^{-2}}{\sum_i w_j^2 s_j^{-2}}$$

and the standard error estimated by

$$SE(\tilde{\alpha}) = \frac{1}{\sqrt{\sum_i w_j^2 s_j^{-2}}}$$

Here $\hat{\beta}_j$ is the regression coefficient when the outcome $y_i$ is regressed onto the *j*th SNP, and $s_j$ is corresponding standard error of $\hat{\beta}_j$. In order to investigate the bi-directional effects of polygenic risk of SCZ and BD on metabolic traits, we performed two sets of analyses, one using PRS from the psychiatric disorders to regress on metabolic traits and the other using PRS from metabolic traits to regress on SCZ or BD. The above formula is equivalent to the "inverse variance weighted" (IVW) approach in MR studies. However, in a polygenic score analysis we do not impose stringent inclusion criteria for genetic variants: we do not require the variants to be strongly associated with the disease and pleiotropic effects are allowed.

*Polygenic risk score testing in the Northern Finland Birth Cohort 1966*
In the second approach, PRS testing was performed on individual genotype and phenotype data in the Northern Finland Birth Cohort (NFBC) 1966. The raw genotype and phenotype data was accessed from dbGaP (accession number phs000276.v2.p1). The original study has passed all relevant ethnical approvals. All study subjects were recruited from Norther Finland and all traits in this study were measured at 31 years old. GWAS has been performed on this cohort using the Infinium 370cnvDuo array [45]. We included a range of metabolic traits in our analyses, including TG, HDL, LDL, BMI, WHR, FG, INS, systolic blood pressure (SBP) and diastolic blood pressure (DBP).

Standard GWAS quality control procedures were performed. Genetic variants with missing rate >10%, minor allele frequency <0.01 and with significant deviation from Hardy-Weinberg equilibrium ($p<10^{-5}$) were excluded. Individuals with genotyping rates <90% were excluded from analyses. We also excluded individuals who are pregnant, receiving anti-diabetic medications, or not fasted before blood glucose and lipid measurements. After the above filtering and quality control procedures, 4982 individuals and 334458 SNPs were retained for further analyses.

Linear regression was carried out to test for the association of PRS with each of the metabolic traits. We included gender and the top ten principal components (PCs) derived from GWAS data as covariates [46]. The analyses were repeated with and without BMI as a covariate.

For both types of analyses (summary statistics and individual genotype-based) listed above, PRS analyses were performed by the R program PRsice [47]. Linkage disequilibrium (LD) clumping was performed prior to polygenic scoring. This procedure preferentially selects markers with lower *p*-value while pruning out variants having a certain degree of LD with the selected ones. We used an $R^2$ threshold of 0.25 for testing PRS associations in the NFBC sample. The method gtx however requires a stronger assumption on the independence between SNPs, and we chose an $R^2$ threshold of 0.05, following the suggestions given in the PRsice vignette. We used the 1000 Genome Project sample as reference LD panel for gtx analyses. Polygenic scores were calculated over a range of *p*-value thresholds (0.001, 0.005, 0.01, 0.03, 0.05, 0.1, 0.2, 0.3, 0.4, 0.5).

Most polygenic score analyses involved quantitative cardiometabolic traits. This approach avoids bias and false positive associations due to potential sample overlap. As shown by Burgess et al.[48], Mendelian randomization studies (using a similar mathematical formulation as gtx) do not suffer from bias if risk factor measurements are made only in the control sample, even in a one-sample setting (i.e. 100% of controls contribute to the risk factor measurement). In our case, we have checked that all continuous cardiometabolic traits were *not* measured in SCZ or bipolar patients; and despite possible overlap in control subjects, the overlap is far less than 100%, hence the chance of false positive associations is low. To further check for sample overlap, we also examined the intercept from LD score regression[49]. We examined whether the intercept is significantly different from zero as an additional method of checking for overlap. (The intercept should be close to zero when there is no overlap in samples)[49]. As will be described in the results section, there is no evidence of significant overlap except for two case-control traits, DM and CAD. We also included another DM GWAS dataset derived from Japanese subjects to ensure the absence of overlap with SCZ/BD samples.

Multiple testing was corrected by the false discovery rate (FDR) procedure, which controls for the expected proportion of false positives[50]. FDR was computed for each pair of traits, and as shown by Efron[51], the overall FDR is also controlled at the same level. The R program p.adjust was used to compute the corresponding FDR for every hypothesis test, which is also known as the *q*-value (the proportion of null was set to be one in all calculations). A *q*-value smaller than 0.05 is regarded as significant while results with *q*-values between 0.05 and 0.1 are considered suggestive associations.

**Cross-trait LD score regression (LDSC)**
Cross-trait LD score regression was performed to assess genetic correlations[49]. We employed the LDSC

program (available at https://github.com/bulik/ldsc) for the analysis, following default parameter settings. This approach allows for sample overlap and accounts for LD between markers. The program "popcorn", which accounts for differing ethnic groups of samples, was used for LD score regression involving the Japanese DM sample[52]. As explained earlier, the significance of the cross-trait LDSC intercept was assessed by dividing the estimated intercept by its standard error. LD score regression was not conducted for SBP and DBP as these were exome-based studies and LDSC was not designed for such studies.

**Mendelian randomization (MR) analysis**

Next, we performed MR analysis to assess causal relationships between SCZ/BD with CM disorders. MR utilized genetic variants as "instruments" to represent the risk factor and analyzed the relationship with an outcome[53,54]. Intuitively, MR is analogous to a randomized controlled trial (RCT). For example, in an RCT of a LDL-lowering drug, patients are randomized to receiving the drug or placebo; similarly, in an MR study, patients with *genetically* lower LDL (as determined by one or multiple variants) are analogous to receiving the lipid-lowering drug. Due to the random allocation of alleles by Mendel's second law, the process is "randomized", and MR is sometimes known as the "nature's RCT". Compared to conventional observational studies, MR are less susceptible to confounding and reverse causality[53,54].

In the current study, we performed two-sample MR with GWAS summary statistics using the R package "MR-base"[55]. SNPs passing genome-wide significance ($p < 5e-8$) were selected as instruments for exposures. The SNPs are first LD-clumped following the default settings in MR-base (clumping window of 10000kb, $R^2$ threshold of 0.01). We conducted MR with three commonly used approaches, including the inverse-variance weighted (IVW)[41,42,56], MR Egger[57] and weighted median methods[58]. The latter two methods can be used to account for horizontal pleiotropy (i.e. the instrument genetic variant(s) affect the outcome via pathways other than through the exposure). The Egger method gives unbiased estimate of the causal effect in the presence of directional pleiotropy, under the assumption that the pleiotropic effects are independent of the genetic association with the risk factor ("InSIDE" assumption)[57]. The weighted median approach gives valid estimates when at least half of the information comes from valid instrumental variables[58]. We primarily reported the IVW estimates, however if Egger's method reveal significant (unbalanced) horizontal pleiotropy, the latter two estimates are reported. We did not report the results for DM[32] and CAD[34] due to potential issues of overlap.

Note that in polygenic score analyses horizontal pleiotropy is allowed but it will affect the validity of casual inference in MR. Polygenic score studies can elucidate shared genetic bases between disorders but is not designed for casual inference.

**Discovering shared SNPs**

For each pair of diseases (psychiatric versus metabolic), we computed for every SNP the probability of being associated with *both* diseases, given the observed test statistics. It can be viewed as a two-dimensional extension of the local true discovery rate [59]. The approach is closely related to the conditional false discovery rate [23] and a similar formulation has been proposed. However, our focus is on the probability of *shared* associations instead of chance of associations *conditioned* on another trait. The methodology we used here followed that proposed by Chung et al. [60], although we worked with the z-statistics instead of *p*-values. Briefly, we assumed a four-group mixture model of *z*-statistics:

$$f(z_A, z_B) = p_{00}f_{00}(z_A, z_B) + p_{10}f_{10}(z_A, z_B) + p_{01}f_{01}(z_A, z_B) + p_{11}f_{11}(z_A, z_B) \quad ----(1)$$

where $z_A$ and $z_B$ refer to the *z*-statistics of trait A and trait B respectively. For a genetic variant, it can be associated with none of the traits, with one trait only, or both. The four possibilities are modeled in the above equation following Chung et al[60]. For instance, $p_{11}$ denotes the proportion of markers that has association with both traits, and $f_{11}$ denotes the probability density function of the z-statistics of this group of markers. From equation (1) we can derive the probability that a SNP is associated with both traits (denoted by $tdr_{11}$), using an expectation-maximization (EM) algorithm. Further details of the method are given in Supplementary Information. We also conducted the same analysis on CAD and DM and presented the results for reference, although sample overlap is present. This is because as the analysis is SNP-based, the effect of sample overlap will not be accumulative across variants as in a PRS analysis (PRS involves adding up the effects of a large number of variants); also, the ranks of shared genetic variants are unlikely to be affected even when there is overlap, since the correlations between samples influence each SNP in similar ways. Moreover, as shown in[48], the bias due to overlap is inversely related with effect size of genetic variants. Since we only focus on the top shared SNPs, any bias due to overlap is likely to be small.

**Pathway analysis**

We selected SNPs with $tdr_{11} >= 0.5$ and mapped those SNPs to genes using the Bioconductor package BioMart using default settings. We noted that some SNPs were mapped to multiple genes which may bias the results of gene set analyses. We adopted the following method to correct for this bias. We first extracted information on variant consequences (using sequence ontology [SO] terms) via BioMart. Each SNP was then mapped to the gene which corresponds to the highest impact rating (as listed in http://asia.ensembl.org/info/genome/variation/predicted_data.html#consequences). For example, for a SNP listed as a frameshift variant in gene A and an intron variant in gene B, the SNP is mapped to gene A only. The pathway analyses based on hypergeometric tests. We performed analyses by ConsensusPathDB, which integrates a variety of databases such as KEGG, Reactome, Wikipathways and BioCarta into a single resource. Analyses were separately carried out for variants having the same, opposite or either directions of associations with SCZ/BD and CM traits.

**Results**

**SCZ and cardiometabolic traits**

The associations of SCZ with cardiometabolic traits using paired summary statistics are shown in Table 1. PRS were first constructed from SCZ GWAS data and CM traits were treated as the outcomes (*i.e.* target phenotypes). Then the analyses were repeated with PRS derived from CM traits. Intuitively, PRS from CM traits can be regarded as a proxy of the actual level of the trait and a previous study [61] has shown that such a polygenic scoring approach can be used to test association of biomarkers with disease. For binary traits, the polygenic scores may be regarded as the underlying liability to the corresponding disorder.

Overall the strongest association was observed for BMI (lowest *p* = 2.82E-12) in polygenic score analyses when SCZ was used either as the predictor or target phenotype. Interestingly, the coefficient was negative, signifying an inverse association. For lipid traits, we observed a positive polygenic association of SCZ with HDL, but no significant associations with LDL and TG. The polygenic association with LDL reached nominal significance (at the optimal p-value threshold), but did not pass the preset FDR threshold of 0.1. For the two adipokines on the list, we observed positive polygenic associations with leptin (BMI-adjusted) and inverse association with adipokine when SCZ PRS was used as the predictor variable. It is worth noting that both raised leptin and reduced adiponectin were associated with heightened CV risks[62,63]. As for measures of obesity and fat deposition (other than BMI), positive association with WHR was observed, especially for BMI-adjusted WHR. Positive association was also seen for BMI-adjusted visceral adiposity, although we observed a negative link of SCZ PRS with subcutaneous adipose tissue mass (unadjusted for BMI). Finally, for traits related to glucose metabolism, PRS of FG was positively associated with SCZ and similar association was observed for PRS derived from fasting insulin. In addition, we observed a positive polygenic link of SCZ PRS with type 2 DM. LD score regression revealed a significant genetic correlation of lower BMI with SCZ, although there were no other significant results. The intercepts from cross-trait LD score regression, which also served as a test for sample overlap, were largely non-significant. The exceptions included CAD and DM, which were expected due to overlap in control subjects. While another trait (adiponectin) also showed p < 0.05 for the LD score intercept, we expect (0.05*26) ~ one "false-positive" at an alpha level of 0.05 (not counting CAD/DM), therefore the results were consistent with the absence of significant sample overlap affecting the results.

For analyses on the Northern Finland Birth Cohort, the most significant association was WHR adjusted for BMI (lowest p = 0.013, q = 0.055) (Supplementary Table 1). Other results were largely non-significant, probably owing to a much smaller sample size of NFBC compared to other GWAS meta-analyses.

In the MR analysis (Table 3), when SCZ was considered as the risk factor, the most significant result was observed for TG. There was evidence of horizontal pleiotropy (p = 0.001), so we focused on results from Egger regression and the weighted median method. The Egger's method suggested that SCZ may be causally related to an increase in TG (beta= 0.186 [95% CI: 0.093 – 0.279], p = 3.37E-4, q = 0.009) while the estimate from the weighted median approach showed a trend towards significance (p = 0.077). We also checked the test results for heterogeneity and found no significant heterogeneity for the MR Egger method (p = 0.208). We observed a nominally significant association with pericardial fat volume by the IVW approach, which was stronger after adjusting for BMI (beta= 0.072 [95% CI: 0.011 – 0.134], p = 0.021, q = 0.288), although both results did not pass the preset FDR threshold. We did not observe significant causal relationship with other

CM traits. When SCZ was regarded as the outcome and CM traits as risk factors, there was no evidence of causal relationships for any CM trait, except that insulin was weakly significant (beta = 1.066 [95% CI: 0.040 – 2.091], p = 0.042) but did not pass FDR correction.

The full results of the shared SNPs analyses with $tdr_{11}$>0.5 are presented in Supplementary Table 5. We also present "gene-based" results for the top genes mapped to the shared SNPs (Supplementary Table 7). The top 3 genes shared between SCZ/BD and each CM trait are shown in Table 5. Across all CM traits, we discovered in total 15422 and 2890 shared genetic loci with $tdr_{11}$> 0.5 and >0.8 respectively (as determined by clumping at an $R^2$ threshold of 0.1 using 250-kb windows; Supplementary Table 11). As ConsensusPathDB outputs enrichment results from numerous pathway databases, there was some overlap with similar pathways and involved genes. Also due to the very long list of pathways involved for different traits, we present selected top pathways in Table 6, taken from the top 25 most commonly top-ranked pathways across all traits (with reduced repetitions of similar pathways). The full results of pathway analyses are presented in Supplementary Table 9. We also presented summary results of pathways derived from SNPs having the same direction of associations with CM disorders, given the polygenic association of SCZ with worse CM risks with regards to several metabolic parameters.

**Bipolar disorder and cardiometabolic traits**

Table 2 shows the polygenic associations of BD with cardiometabolic traits using paired summary statistics. Similar to SCZ, we observed an inverse association with BMI ($p$ = 9.34E-13) when BD was treated as the target or base phenotype. We also observed a polygenic association of lower WHR with BD ($p$ = 1.35E-03), but the result became non-significant after adjustment by BMI, suggesting the original association is mainly driven by BMI. For lipid traits, we observed polygenic associations of BD with higher HDL, as well as lower LDL, TC and TG. We also detected polygenic associations with lower leptin levels and SBP.

In LD score regression, none of the genetic correlations achieved q-value < 0.1 but a number of traits were nominally significant with q < 0.2. These included BMI and WHR, which showed negative genetic correlations with BD, consistent with results from the PRS analysis. Fasting glucose and insulin showed nominally significant negative genetic correlations with BD. Interestingly, we observed a marginally significant positive correlation with DM. The overall direction of genetic correlations from LD score analysis leaned towards lower CM risks, broadly in line with PRS analyses. The LD score regression intercepts were again mostly non-significant apart from CAD and DM.

As for the analyses of the NFBC cohort, we did not observe any significant results passing FDR correction (Supplementary Table 2). Again the absence of significant associations may be due to the relatively small sample size of NFBC cohort compared to other GWAS meta-analyses. Nevertheless, for a few traits showing at least p<0.05 at the optimal p-value threshold (including INS, TG and SBP), the directions of associations were negative, largely in line with the summary PRS and LD score analyses.

In the MR analysis (Table 4), when BD was treated as the exposure, there was no evidence of casual relationship with CM traits. On the other hand, when CM traits were considered as exposures, we observed several nominally significant results. For example, lower FG appeared to be casually related to BD (beta = -4.474, SE = 0.183, p = 0.01). Results for SBP and WHR (BMI adjusted) were only nominally significant but did not pass FDR correction. The association with BMI was statistically significant based on Egger's approach, but there was significant heterogeneity among the casual effects of individual variants ($p = 0.022$). This may reflect the genetic instruments were measuring different quantities, casting doubt on the assumption of MR being valid for all genetic variants. We did not observe evidence of heterogeneity for other nominally significant results.

The results of the shared SNPs analyses are presented in Supplementary Tables 6 and 8, the latter showing genes mapped to the top shared SNPs. Across all CM traits, we found totally 3115 and 297 shared genetic loci with $tdr_{11}$> 0.5 and >0.8 respectively (after clumping the original results at an $R^2$ threshold of 0.1; see Supplementary Table 11). As GWAS sample size for BD is smaller, the number of shared SNPs with high $tdr_{11}$ was smaller. Selected top pathways are given in Table 6 while Supplementary Table 10 shows the full results of gene-set enrichment analysis.

**Discussion**

In the current study, we have performed polygenic risk score analyses on the possible shared genetic basis of SCZ and BD with a comprehensive panel of cardiometabolic traits. We observed a number of potential polygenic associations, which will be discussed below.

*SCZ and cardiometabolic traits*

Raised cardiovascular morbidity and mortality is well-established in SCZ, yet it is difficult to disentangle the numerous possible underlying factors, including the side-effects of antipsychotics. Broadly in line with previous studies, we observed here a positive polygenic association of higher FG with SCZ. Similarly, a positive polygenic link of SCZ with DM was detected. It is notable that multiple studies have found impaired glucose tolerance or insulin resistance in drug-naïve patients and their first-degree relatives [13-16,64-69].

Despite increased cardiovascular risks in SCZ patients, surprisingly, we found evidence for an inverse relationship between polygenic scores of BMI and SCZ. Interestingly, a population cohort study of over one million men in Sweden also reported that subjects with lower BMI had an increased risk of developing SCZ [70]. A similar study in Denmark reached the same conclusions [71]. A few studies in drug-naïve patients also revealed a trend of lower BMI. For instance, lower BMI among drug-free SCZ patients was reported by Spelman et al. [13] and Pandmavati et al. [72]. However, some studies failed to detect any differences in BMI between SCZ cases and controls [64,65]. The underlying pathophysiology for an inverse relationship between BMI and SCZ is unknown. One hypothesis is that poor nutritional status, however subtle, may adversely affect neural development [70] which leads to psychotic disorders.

While we observed that lower BMI polygenic score is associated with SCZ and vice versa, we observed the opposite trend for WHR. Polygenic association of higher WHR with SCZ, especially after adjustment for BMI, is shown both in the Northern Finland cohort sample and analyses using summary statistics. Moreover, this study revealed potential polygenic link of SCZ with increased visceral adiposity, and an unfavorable profile of adipokines (raised leptin and decreased adiponectin). Consistent with this finding, Ryan et al. reported higher WHR and three times more intra-abdominal fat in drug-free SCZ patients compared to BMI-matched controls [73]. Others[74,75] also reported increased WHR in drug-free patients compared to controls. While some studies did not find any differences in WHR[64,76], they were of small sample sizes and inadequate power may explain the inconsistent results. Waist-hip ratio reflects visceral fat deposition, and has been suggested to be an informative predictor of cardiovascular risks independent of BMI[77,78]. As for adipokine abnormalities, a recent meta-analyses reported elevated leptin levels in SCZ patients, although the increase was moderate[79]. However, many studies in the meta-analyses involved patients taking antipsychotics, and for those studies on antipsychotic-naïve patients, the sample sizes were small. With regards to adiponectin, meta-analysis showed that SCZ patients taking second-generation antipsychotics (SGAs), especially clozapine and olanzapine, had significantly lower adiponectin levels[80]. Here we showed that SCZ patients may be genetically associated with adverse adipokine profiles (which are linked to raised CV risks), independent of antipsychotic effects.

For lipid traits, we observed polygenic associations with higher HDL levels. The associations with LDL and TC were not significant but the directions were negative (and nominal significance was reached if only the best p-value threshold was considered), consistent with a favorable cholesterol profile. However, in the MR analysis, SCZ appeared to be causally associated with increased TG. These findings were generally consistent with a very recent meta-analysis[17] on lipid profiles in first-episode psychosis patients which reported higher TG but favorable cholesterol profiles among patients, although they did not reveal changes in HDL.

In general, we did not find evidence for causal relationship between SCZ and CM traits, with the exception of TG. This suggests that SCZ per se may not directly cause changes in CM traits; the associations of SCZ (as a disease itself, without considering medication side-effects) with CM abnormalities are more likely due to *shared* genetic or environmental risk factors. In other words, similar risk factors may influence the risks of *both* SCZ and CM disorders together, but through different pathways. The shared genetic factors are addressed in this study but some previous studies also suggested shared environmental factors for the two kinds of disorders. For example, both SCZ and DM may be associated with risk factors during early development, such as low birth weight and maternal malnutrition[81,82]. For example, individuals conceived or in early gestation during famines are more likely to develop both SCZ and type 2 DM[83,84].

***Bipolar disorder and cardiometabolic traits***
Similar to SCZ, we observed an inverse genetic relationship between BMI and BD. One previous clinical study reported a higher prevalence of being overweight among bipolar patients [85]. However, the weight measurements were not made during euthymic period and 71.4% of the patients were depressed during the measurement. It is possible that the symptoms of bipolar depression contributed to the findings. As remarked by the authors, their findings might be more related to the current episode of illness than to the underlying disorder [85]. Also the comparison group was patients with obsessive-compulsive disorder, not healthy population controls. While WHR was also inversely related to BD in the unadjusted analysis, the relationship no longer exists after controlling for BMI.

There are very few studies on the metabolic parameters in drug-naive bipolar patients in comparison with healthy controls. We do not observe evidence of polygenic association of hyperlipidemia with BD; on the contrary, there appeared to be an inverse relationship with LDL, TC and TG. This result could be confounded by BMI, although studies have found lower cholesterol level in bipolar patients [86,87]. There is also evidence of a correlation between low cholesterol levels and suicide risks [88]. The underlying mechanism remains elusive, but it has been suggested that low cholesterol may be associated with decreased lipid microviscosity in neural membranes and reduced exposure of serotonin receptors on membrane surface. Serotonergic dysfunction may in turn contribute to suicidality [89]. Overall we observed polygenic associations of BD with favorable metabolic parameters, suggesting that other secondary risk factors (e.g. medications) may play a larger part than the disorder itself in affecting cardiovascular risks. One exception is that the genetic correlation with DM was positive, although the correlations with fasting glucose and insulin were negative. One possibility is that some findings may represent false positives, given that they were only nominally significant and did not pass our preset q-value threshold. Another explanation is that the study on fasting glucose and insulin was performed on non-diabetic subjects[30], and genetic factors associated with normal-range glucose levels may not fully overlap with the variants conferring susceptibility to frank diabetes[90,91].

As for the results from MR analysis, there was no evidence of causal relationships when BD was treated as the exposure. When BD was considered as the outcome, lower FG appeared to be casually linked to BD. We are not clear about the underlying mechanisms, but interestingly a recent study reported greater mood lability being associated with lower fasting glucose levels[92]. However, yet another small-scale study patients revealed more glucose abnormalities in drug-naïve bipolar patients[18]. Further clinical studies are required to elucidate the exact relationship. Other MR results did not pass FDR correction, and for BMI there was high heterogeneity which cast doubt on the validity of the instruments.

**Clinical implications**
Our findings provide evidence to support increased awareness and better monitoring and management of CV risks in psychosis patients, especially for SCZ. Taken together, we found evidence of polygenic associations with glucose metabolism abnormalities, adverse adipokine profiles, central obesity and increased visceral adiposity in SCZ. These findings are grossly in line with observations from epidemiological studies. If our findings are validated in further studies, given that SCZ patients may be associated with several CM abnormalities independent of medication side-effects, proper surveillance and management of CV risk factors may be required from the onset of the disease, and even in patients who are taking drugs with less metabolic side-effects or low-dose medications. Our findings also suggest that the measurement of WHR (instead of BMI alone) may be useful in screening for metabolic risks, due to its closer association with visceral adiposity. Also, adipokines such as leptin and adiponectin may warrant further investigations as biomarkers for CV risks in SCZ patients.

On the other hand, we observed polygenic associations with favorable cholesterol profiles in SCZ and BD, and an overall favorable CM profile in BD patients. These results suggest that hypercholesterolemia in SCZ and many CM abnormalities in BD may be secondary (and hence more readily modifiable) instead of being related to the underlying pathophysiology of SCZ/BD. One of the most important risk factor would be side-effects from antipsychotics and mood stabilizers; careful choice and dosing of these drugs will likely reduce CV risks in such patients. Promotion of healthy lifestyle (e.g. proper diet and exercise) might also be beneficial to SCZ/BD patients.

*Shared SNP analyses of SCZ and bipolar disorder with cardiometabolic traits*

We identified a number of susceptibility variants that are potentially shared between SCZ or BD and cardiometabolic traits. As the list is long, we will not discuss each susceptibility gene in detail but will highlight a few associated pathways here. For example, it is interesting to note that pathways associated with immune functioning (*e.g.* antigen processing, complement pathways) were ranked highly among a number of cardiometabolic traits, such as HDL, LDL, TG, DBP and WHR, and such pathways were ranked high in both SCZ and BD. If we limit our analyses to shared SNPs with the same directions of effects on SCZ and CM risks, the pathway most frequently listed was antigen processing and presentation, which was listed as a significant pathway in 7 metabolic traits. Immune system dysfunction has been implicated in SCZ, BD [93] as well as cardiometabolic traits [94]. Recent studies have suggested that chronic inflammation may be an important mediator linking metabolic abnormalities and severe mental illnesses [95]. For example, elevations of pro-inflammatory cytokines such as tumor necrosis factor alpha and interleukin-6 have been observed both in patients with psychosis and metabolic syndrome. Chronic systemic inflammation may be coupled with microglia activation in the brain, which may disturb neuronal functions [95]. On the other hand, inflammation occurring in visceral adipose tissues, the liver, pancreatic islets and blood vessels contribute to metabolic abnormalities and atherosclerosis [96]. Our findings support a potential role of inflammation and immune system pathways in the shared genetic bases of SCZ and BD with cardiometabolic abnormalities. In addition, we also observed that lipid metabolism and the statin pathway were among the pathways most often implicated. Statins have been known to lower the risk of coronary artery disease [97], yet it has also been investigated as adjunctive therapy for SCZ [98]. Vincenzi et al. [98] observed a significant decrease in positive symptoms of SCZ patients from baseline to 6 weeks with pravastatin treatment, though the decrease was not significant at 12 weeks. It was postulated that the anti-inflammatory actions of statin contributed to the therapeutic effect on SCZ. It is worthwhile to note that the anti-inflammatory potential of statins has been postulated as a chief mechanism responsible for preventing CAD [99] as well. Another pathway listed was EGFR-1 signaling pathway, which was implicated in multiple facets of cardiovascular abnormalities, such as endothelial dysfunction and atherosclerosis (reviewed in ref.[100]). The EGF family and related proteins such as ErbB1-4 also serve important functions in brain functions and pathology of SCZ[101].

*Study limitations*
There are a few limitations to our study. Our analyses are based on large-scale GWAS meta-analyses results, but most studies were done in Caucasians. It is worthwhile to extend the study to other ethnic groups. For analyses using paired summary results, we were unable to control for other clinical factors, for instance adjusting for WHR or BMI when studying lipid traits. Further clinical studies of metabolic abnormalities in SCZ and bipolar patients, coupled with genetic testing, will be useful in delineating the susceptibility markers and providing more solid evidence of shared genetic susceptibilities. We have employed two different approaches, PRS analyses and LD score regression, to assess shared genetic bases between SCZ/BD and CM traits. However, each method has its limitations. For instance, it is relatively difficult to account for LD and sample overlap in PRS analyses. Although LD score regression takes into account these factors, it relies on other assumptions, for example it assumes that the casual variants are randomly distributed in the genome regardless of the LD structure, and that (ideally) the SNP effects are of equal variances (i.e. rare SNPs should have larger effect sizes and vice versa, such that the variance explained should be the same for each SNP)[49,102]. Deviation of effect sizes from normality may also render the procedure inefficient[49]. While it was shown that these assumptions may not influence the results much under moderate violations, it is unclear how these assumptions may fit for different diseases with diverse genetic architecture. The difference in the underlying methodology and assumptions may explain why PRS and LDSC results sometimes do not agree with each other. In addition, both methods as well as MR assume linearity of SNP effects. Also, SCZ and BD are likely to be heterogeneous disorders, and the association with CM abnormalities may differ within different patient subgroups.

While we have discussed a few pathways that may be involved in the shared pathophysiology, further experimental studies are required to elucidate the exact mechanisms involved. Many other shared SNPs and pathways also warrant further investigations. Interestingly, we did not find uniformly increased polygenic risks of all metabolic abnormalities in SCZ and BD. Further investigations into the complex links between CVD risk factors and SCZ and BD might shed light on new therapeutic measures for both types of disorders.

**Figure legends**
Figure 1:    The overall analytic workflow. First, we made use of large-scale GWAS meta-analyses results for SCZ, BD and a comprehensive panel of 28 metabolic and cardiovascular traits to test for shared genetic bases by polygenic risk scores (PRS) and LD score regression. We also examined PRS associations with cardiovascular risk factors in the Northern Finland Birth Cohort with individual genotype and phenotype data.

We then performed MR analysis to assess causal relationship between the two groups of disorders. Finally, we "zoomed in" to discover the genetic variants shared between SCZ and BD with each metabolic trait, and inferred the likely involved pathways by gene-set analyses.

**Acknowledgements**

This study was partially supported by the Lo Kwee Seong Biomedical Research Fund and a CUHK Direct Grant to H-.C.S. The authors declare no conflicts of interest. We would like to thank Prof. Stephen Tsui and the Hong Kong Bioinformatics Centre for bioinformatics support.

**Author contributions**

Conceived and designed the study: HCS. Performed core data analyses: HCS and CKLC. Provided advice on computational/statistical analyses: PCS. Performed pathway analysis: F-K.A, C-H.M.
Drafted the manuscript: HCS. Supervised the study: HCS.

Supplementary files are available at
https://drive.google.com/open?id=1zue7QPftcoqzJUHZ9nqLccroSf4ibEzh

Table 1 Polygenic association testing of SCZ with cardiometabolic traits using summary statistics

| | Polygenic score analysis | | | | | | | | LD Score | | |
|---|---|---|---|---|---|---|---|---|---|---|---|
| | SCZ as target | | | | SCZ as base | | | | | | |
| | best_p | pval | coef | qvalue | best_p | pval | coef | qvalue | rg | p | qval |
| Adiponectin | 0.05 | 7.87E-02 | -4.82E-03 | 3.97E-01 | 0.4 | 1.62E-03 | -3.03E-03 | **6.92E-03** | -0.002 | 0.967 | 0.968 |
| BMI | 0.05 | 2.82E-12 | -6.18E-02 | **1.67E-11** | 0.03 | 1.22E-10 | -6.40E-03 | **1.22E-09** | -0.068 | **0.006** | 0.157 |
| CAD | | | | | | | | | -0.025 | 0.375 | 0.828 |
| DBP | 0.03 | 2.42E-01 | 2.62E-03 | 8.56E-01 | 0.5 | 1.32E-01 | -3.44E-02 | 5.60E-01 | | | |
| DM | | | | | | | | | -0.033 | 0.387 | 0.828 |
| DM-2$^{nd}$set | 0.3 | 7.74E-02 | 3.00E-03 | 3.43E-01 | 0.5 | 2.99E-03 | 2.03E-02 | **2.13E-02** | 0.067 | 0.096 | 0.705 |
| Fat-percentage | 0.2 | 1.65E-02 | 4.42E-03 | *9.52E-02* | 0.1 | 1.93E-01 | 1.65E-03 | 6.29E-01 | -0.016 | 0.601 | 0.843 |
| FG | 0.05 | 1.28E-02 | 1.11E-02 | **4.67E-02** | 0.001 | 1.17E-01 | -2.21E-03 | 3.16E-01 | -0.029 | 0.383 | 0.828 |
| FG-adjBMI | 0.005 | 7.48E-02 | 1.53E-02 | 2.63E-01 | 0.001 | 7.85E-02 | -2.55E-03 | 3.57E-01 | -0.021 | 0.537 | 0.843 |
| HDL | 0.2 | 8.83E-04 | 1.74E-02 | **3.36E-03** | 0.01 | 7.23E-07 | 7.26E-03 | **6.27E-06** | 0.046 | 0.165 | 0.705 |
| INS | 0.005 | 5.50E-03 | 2.06E-02 | *5.50E-02* | 0.03 | 5.41E-02 | -1.65E-03 | 5.41E-01 | 0.013 | 0.789 | 0.888 |
| INS-adjBMI | 0.005 | 3.64E-02 | 1.73E-02 | 3.56E-01 | 0.03 | 4.84E-02 | -1.44E-03 | 4.68E-01 | 0.015 | 0.783 | 0.888 |
| LDL | 0.1 | 1.26E-02 | -1.33E-02 | 1.26E-01 | 0.05 | 4.73E-02 | -2.54E-03 | 1.81E-01 | -0.025 | 0.442 | 0.828 |
| Leptin | 0.001 | 6.92E-03 | 4.09E-02 | *6.92E-02* | 0.1 | 2.86E-02 | 3.27E-03 | 1.55E-01 | 0.069 | 0.140 | 0.705 |
| Leptin-adjBMI | 0.5 | 1.95E-02 | 1.03E-02 | 1.83E-01 | 0.005 | 5.86E-03 | 4.80E-03 | **3.28E-02** | 0.092 | 0.060 | 0.705 |
| PAT | 0.5 | 3.55E-02 | -3.34E-03 | 1.25E-01 | 0.05 | 4.03E-01 | -2.74E-03 | 8.99E-01 | 0.053 | 0.460 | 0.828 |
| PAT-adjHtWt | 0.4 | 5.42E-01 | -9.70E-04 | 9.47E-01 | 0.001 | 1.84E-01 | 7.88E-03 | 9.64E-01 | 0.054 | 0.352 | 0.828 |
| SAT | 0.03 | 4.02E-02 | -5.11E-03 | 2.27E-01 | 0.03 | 5.84E-03 | -7.60E-03 | **3.72E-02** | -0.026 | 0.596 | 0.843 |
| SATHU | 0.2 | 2.92E-01 | -2.06E-03 | 8.96E-01 | 0.5 | 4.55E-01 | -1.89E-03 | 9.87E-01 | -0.210 | 0.209 | 0.705 |
| SBP | 0.2 | 9.95E-02 | -1.15E-03 | 6.04E-01 | 0.01 | 7.05E-03 | -1.59E-01 | *7.05E-02* | | | |
| TC | 0.3 | 4.27E-02 | -9.38E-03 | 1.99E-01 | 0.03 | 2.92E-01 | -1.40E-03 | 8.39E-01 | -0.023 | 0.422 | 0.828 |
| TG | 0.3 | 8.80E-02 | -8.71E-03 | 3.33E-01 | 0.001 | 1.83E-01 | 2.70E-03 | 9.76E-01 | -0.043 | 0.142 | 0.705 |
| VAT | 0.001 | 2.64E-02 | 1.56E-02 | 2.64E-01 | 0.005 | 3.47E-02 | -7.60E-03 | 3.47E-01 | 0.005 | 0.934 | 0.968 |
| VAT-adjBMI | 0.1 | 1.12E-03 | 6.50E-03 | **1.12E-02** | 0.5 | 5.50E-03 | 5.94E-03 | **3.52E-02** | 0.027 | 0.625 | 0.843 |
| VATHU | 0.001 | 1.08E-02 | -2.01E-02 | 1.08E-01 | 0.5 | 2.86E-01 | -2.68E-03 | 7.79E-01 | -0.030 | 0.670 | 0.862 |
| VATSAT | 0.005 | 2.66E-01 | 4.58E-03 | 9.08E-01 | 0.005 | 1.39E-01 | -5.35E-03 | 7.29E-01 | 0.033 | 0.549 | 0.843 |
| VATSAT-adjBMI | 0.1 | 5.12E-01 | -1.33E-03 | 8.43E-01 | 0.005 | 1.54E-01 | -5.15E-03 | 9.42E-01 | 0.019 | 0.744 | 0.888 |
| WHR | 0.5 | 3.75E-02 | 1.10E-02 | 3.43E-01 | 0.3 | 2.87E-02 | 1.95E-03 | 1.02E-01 | -0.032 | 0.203 | 0.705 |
| WHR-adjBMI | 0.2 | 1.91E-04 | 2.11E-02 | **1.50E-03** | 0.5 | 3.91E-05 | 3.63E-03 | **1.92E-04** | -0.001 | 0.968 | 0.968 |

SCZ was treated as the target phenotype (i.e. dependent variable) on the left block and treated as the base phenotype (i.e. as predictor variable) in the middle block. Best *p*, best p-value threshold; coef, regression coefficient; rg, genetic correlation.

BMI, body mass index; CAD, coronary artery disease; DM, type 2 diabetes mellitus (sample mainly from Caucasians); DM-2$^{nd}$ set, type 2 diabetes (sample from Japanese); FG, fasting glucose; FG.adjBMI, fasting glucose adjusted for BMI; INS, fasting insulin; INS.adjBMI, fasting insulin adjusted for BMI; HDL, high-density lipoprotein; LDL, low-density lipoprotein; TG, triglycerides; TC, total cholesterol; WHR, waist-hip ratio; WHR.adjBMI, waist-hip ratio adjusted for BMI; SBP and DBP, systolic and diastolic blood pressure; SAT, subcutaneous adipose tissue volume; VAT, visceral adipose tissue volume; PAT, pericardial fat volume; SATHU, subcutaneous adipose tissue attenuation; VATHU, visceral adipose tissue attenuation; adjHtWt, adjusted for height and weight.

For PRS analyses, q-values < 0.05 are in bold while suggestive associations (0.05 <= q <= 0.1) are italics. For LD score regression, results with nominal significance (p < 0.05) are highlighted in bold.

Table 2  Polygenic association testing of bipolar disorder with cardiometabolic traits using summary statistics

|  | BD as target | | | | BD as base | | | | LD score | | |
|---|---|---|---|---|---|---|---|---|---|---|---|
|  | best_p | pval | coef | qvalue | best_p | pval | coef | qvalue | rg | p | qval |
| Adiponectin | 0.005 | 1.05E-01 | -1.78E-02 | 3.90E-01 | 0.2 | 5.92E-02 | -1.01E-03 | 2.16E-01 | -0.026 | 0.741 | 0.770 |
| BMI | 0.2 | 3.54E-11 | -9.23E-02 | **3.54E-10** | 0.2 | 7.30E-06 | -1.96E-03 | **3.47E-05** | -0.094 | **0.014** | 0.127 |
| CAD |  |  |  |  |  |  |  |  | 0.017 | 0.652 | 0.765 |
| DBP | 0.4 | 6.36E-01 | -5.01E-04 | 8.78E-01 | 0.2 | 4.00E-02 | -2.82E-02 | 4.00E-01 |  |  |  |
| DM |  |  |  |  |  |  |  |  | 0.112 | **0.043** | 0.151 |
| DM-2$^{nd}$set | 0.4 | 3.06E-02 | -7.63E-03 | 1.61E-01 | 0.5 | 2.40E-02 | -7.34E-03 | 1.41E-01 | -0.053 | 0.415 | 0.717 |
| Fat-percentage | 0.001 | 1.50E-02 | -2.60E-02 | 1.50E-01 | 0.3 | 5.99E-03 | -1.68E-03 | **2.66E-02** | -0.076 | 0.147 | 0.362 |
| FG | 0.001 | 4.14E-02 | -6.00E-02 | 4.14E-01 | 0.005 | 5.59E-02 | -1.55E-03 | 5.59E-01 | -0.121 | **0.044** | 0.151 |
| FG-adjBMI | 0.005 | 1.04E-01 | -2.87E-02 | 7.08E-01 | 0.005 | 1.52E-02 | -2.02E-03 | 1.28E-01 | -0.140 | **0.015** | 0.127 |
| HDL | 0.03 | 2.51E-06 | 7.64E-02 | **2.51E-05** | 0.3 | 2.01E-09 | 3.22E-03 | **2.01E-08** | 0.091 | 0.084 | 0.226 |
| INS | 0.005 | 2.16E-02 | 3.62E-02 | 2.16E-01 | 0.2 | 3.17E-02 | -8.07E-04 | 1.16E-01 | -0.152 | **0.039** | 0.151 |
| INS-adjBMI | 0.3 | 2.95E-02 | -1.63E-02 | 1.03E-01 | 0.01 | 2.49E-02 | -1.31E-03 | 1.20E-01 | -0.155 | **0.019** | 0.127 |
| LDL | 0.4 | 8.94E-05 | -3.74E-02 | **5.41E-04** | 0.3 | 5.49E-03 | -1.62E-03 | **2.53E-02** | -0.020 | 0.708 | 0.765 |
| Leptin | 0.03 | 1.37E-02 | -3.51E-02 | 1.37E-01 | 0.05 | 3.94E-03 | -2.11E-03 | **2.24E-02** | -0.047 | 0.540 | 0.717 |
| Leptin-adjBMI | 0.03 | 3.13E-02 | -2.38E-02 | 1.08E-01 | 0.5 | 1.50E-03 | -2.24E-03 | **6.43E-03** | -0.032 | 0.685 | 0.765 |
| PAT | 0.01 | 5.62E-02 | 1.27E-02 | 3.38E-01 | 0.01 | 2.19E-01 | -3.40E-03 | 9.86E-01 | 0.102 | 0.364 | 0.717 |
| PAT-adjHtWt | 0.005 | 1.63E-01 | -1.11E-02 | 4.87E-01 | 0.01 | 3.31E-01 | -2.70E-03 | 9.96E-01 | 0.056 | 0.553 | 0.717 |
| SAT | 0.05 | 2.22E-01 | 5.90E-03 | 7.71E-01 | 0.005 | 4.15E-01 | -2.17E-03 | 9.99E-01 | 0.018 | 0.799 | 0.799 |
| SATHU | 0.01 | 2.46E-01 | -9.03E-03 | 9.35E-01 | 0.001 | 4.72E-02 | 1.04E-02 | 2.98E-01 | -0.130 | 0.395 | 0.717 |
| SBP | 0.5 | 4.06E-01 | -7.15E-04 | 9.41E-01 | 0.2 | 8.63E-05 | -8.72E-02 | **8.63E-04** |  |  |  |
| TC | 0.3 | 1.02E-02 | -2.53E-02 | **3.27E-02** | 0.05 | 1.83E-02 | -1.74E-03 | 1.83E-01 | -0.108 | **0.045** | 0.151 |
| TG | 0.4 | 8.53E-03 | -2.79E-02 | **3.96E-02** | 0.05 | 2.47E-04 | -2.50E-03 | **2.33E-03** | -0.029 | 0.554 | 0.717 |
| VAT | 0.5 | 1.03E-01 | -6.22E-03 | 3.44E-01 | 0.2 | 2.38E-01 | -1.40E-03 | 7.38E-01 | -0.048 | 0.551 | 0.717 |
| VAT-adjBMI | 0.001 | 2.33E-02 | -3.38E-02 | 2.33E-01 | 0.2 | 6.04E-02 | -2.23E-03 | 2.72E-01 | -0.084 | 0.285 | 0.641 |
| VATHU | 0.5 | 3.44E-01 | 3.72E-03 | 9.39E-01 | 0.2 | 3.58E-01 | 1.27E-03 | 8.33E-01 | 0.081 | 0.431 | 0.717 |
| VATSAT | 0.5 | 1.55E-01 | -5.38E-03 | 5.14E-01 | 0.5 | 1.13E-01 | -1.76E-03 | 3.30E-01 | -0.034 | 0.672 | 0.765 |
| VATSAT-adjBMI | 0.5 | 2.48E-02 | -8.48E-03 | 1.60E-01 | 0.5 | 1.11E-01 | -1.77E-03 | 3.94E-01 | -0.048 | 0.558 | 0.717 |
| WHR | 0.001 | 6.34E-04 | -1.64E-01 | **6.34E-03** | 0.001 | 2.72E-03 | -5.28E-03 | **2.21E-02** | -0.113 | **0.005** | 0.127 |
| WHR-adjBMI | 0.005 | 5.49E-02 | -5.82E-02 | 4.32E-01 | 0.001 | 1.63E-01 | -2.48E-03 | 7.93E-01 | -0.078 | 0.068 | 0.205 |

Bipolar disorder was treated as the target phenotype (i.e. dependent variable) on the left block and treated as the base phenotype (i.e. as predictor variable) in the middle block. Best *p*, best p-value threshold; coef, regression coefficient. Please refer to the legends of Table 1 for other abbreviations.

Table 3    Mendelian randomization analysis of SCZ with cardiometabolic traits

| | SCZ as exposure | | | | | SCZ as outcome | | | |
|---|---|---|---|---|---|---|---|---|---|
| | b | se | pval | qval | | b | se | pval | qval |
| Adiponectin | 0.001 | 0.013 | 0.964 | 0.970 | Adiponectin | 0.008 | 0.108 | 0.941 | 0.941 |
| BMI | -0.019 | 0.013 | 0.148 | 0.691 | BMI | 0.044 | 0.111 | 0.694 | 0.941 |
| DBP | -0.179 | 0.457 | 0.696 | 0.970 | DBP | 0.007 | 0.019 | 0.725 | 0.941 |
| Fat-percentage | -0.002 | 0.013 | 0.893 | 0.970 | Fat-percentage | 0.207 | 0.121 | 0.088 | 0.475 |
| FG | -0.001 | 0.009 | 0.877 | 0.970 | FG | -0.049 | 0.116 | 0.674 | 0.941 |
| FG-adjBMI | 0.000 | 0.009 | 0.970 | 0.970 | FG-adjBMI | -0.191 | 0.106 | 0.071 | 0.475 |
| HDL | 0.023 | 0.024 | 0.332 | 0.970 | HDL | 0.031 | 0.041 | 0.458 | 0.941 |
| INS | 0.004 | 0.007 | 0.597 | 0.970 | INS | 1.066 | 0.523 | **0.042** | 0.475 |
| INS-adjBMI | 0.006 | 0.007 | 0.416 | 0.970 | INS-adjBMI | 0.131 | 0.423 | 0.756 | 0.941 |
| DM-2$^{nd}$set | -0.062 | 0.104 | 0.553 | 0.970 | DM-2$^{nd}$set | -0.009 | 0.022 | 0.678 | 0.941 |
| LDL | 0.008 | 0.013 | 0.534 | 0.970 | LDL | 0.005 | 0.028 | 0.871 | 0.941 |
| Leptin | -0.016 | 0.015 | 0.304 | 0.970 | PAT | -0.029 | 0.084 | 0.732 | 0.941 |
| Leptin-adjBMI | -0.005 | 0.012 | 0.685 | 0.970 | PAT-adjHtWt | -0.059 | 0.060 | 0.328 | 0.941 |
| PAT | 0.060 | 0.030 | **0.047** | 0.443 | SBP | -0.002 | 0.013 | 0.902 | 0.941 |
| PAT-adjHtWt | 0.072 | 0.031 | **0.021** | 0.288 | TC | -0.003 | 0.031 | 0.921 | 0.941 |
| SAT | 0.002 | 0.023 | 0.925 | 0.970 | TG | 0.017 | 0.043 | 0.695 | 0.941 |
| SATHU | 0.011 | 0.029 | 0.704 | 0.970 | VATSAT | 0.264 | 0.158 | 0.095 | 0.475 |
| SBP | 0.259 | 0.605 | 0.669 | 0.970 | VATSAT-adjBMI | 0.163 | 0.135 | 0.230 | 0.920 |
| TC | 0.023 | 0.015 | 0.114 | 0.637 | WHR | -0.008 | 0.104 | 0.940 | 0.941 |
| TG* | 0.186 | 0.048 | **3.37E-04** | **0.009** | WHR-adjBMI | -0.032 | 0.100 | 0.752 | 0.941 |
| VAT | -0.015 | 0.023 | 0.528 | 0.970 | | | | | |
| VAT-adjBMI | -0.004 | 0.023 | 0.850 | 0.970 | | | | | |
| VATHU | 0.002 | 0.030 | 0.946 | 0.970 | | | | | |
| VATSAT | -0.005 | 0.023 | 0.837 | 0.970 | | | | | |
| VATSAT-adjBMI | -0.001 | 0.023 | 0.956 | 0.970 | | | | | |
| WHR | 0.007 | 0.013 | 0.605 | 0.970 | | | | | |
| WHR-adjBMI | 0.016 | 0.014 | 0.255 | 0.970 | | | | | |

b: coefficient estimate from Mendelian randomization; se, standard error; pval, p-value; qval, q-value.

*MR results for TG was taken from Egger's method as there was significant (unbalanced) horizontal pleiotropy (p = 0.001). Several traits do not have SNPs passing genome-wide significance which also overlap with the SCZ GWAS SNPs, and hence are excluded here. Due to sample overlap between CAD and DM (Caucasian sample) with SCZ/BD, these traits are not included in the MR analysis here. P-values and q-values < 0.05 are in bold.

Table 4    Mendelian randomization analysis of bipolar disorder with cardiometabolic traits

| | BD as exposure | | | | | BD as outcome | | | |
|---|---|---|---|---|---|---|---|---|---|
| | b | se | pval | qval | | b | se | pval | qval |
| Adiponectin | 0.015 | 0.013 | 0.252 | 0.913 | Adiponectin | -0.115 | 0.164 | 0.483 | 0.610 |
| BMI | 0.007 | 0.023 | 0.748 | 0.913 | BMI*^ | 0.984 | 0.374 | **0.010** | 0.067 |
| Fat-percentage | 0.019 | 0.024 | 0.410 | 0.913 | DBP | 0.044 | 0.023 | 0.054 | 0.148 |
| FG | -0.004 | 0.009 | 0.638 | 0.913 | Fat-percentage | 0.479 | 0.253 | 0.059 | 0.148 |
| FG-adjBMI | -0.007 | 0.009 | 0.454 | 0.913 | FG | -0.474 | 0.183 | **0.010** | 0.067 |
| HDL | 0.009 | 0.016 | 0.585 | 0.913 | FG-adjBMI | -0.505 | 0.174 | **0.004** | 0.067 |
| INS | -0.003 | 0.009 | 0.745 | 0.913 | HDL | 0.071 | 0.064 | 0.266 | 0.443 |
| INS-adjBMI | -0.009 | 0.008 | 0.257 | 0.913 | INS | -0.997 | 0.733 | 0.174 | 0.316 |
| LDL | -0.017 | 0.018 | 0.367 | 0.913 | INS-adjBMI | -0.572 | 0.597 | 0.338 | 0.520 |
| Leptin | -0.004 | 0.030 | 0.902 | 0.980 | DM-2$^{nd}$set | -0.051 | 0.034 | 0.133 | 0.266 |
| Leptin-adjBMI | 0.000 | 0.026 | 0.994 | 0.994 | LDL | -0.001 | 0.057 | 0.980 | 0.980 |
| PAT | -0.010 | 0.039 | 0.789 | 0.913 | PAT | -0.065 | 0.174 | 0.710 | 0.747 |
| PAT-adjHtWt | -0.028 | 0.039 | 0.476 | 0.913 | PAT-adjHtWt | -0.077 | 0.127 | 0.542 | 0.610 |
| SAT | 0.036 | 0.030 | 0.232 | 0.913 | SBP | 0.024 | 0.010 | **0.021** | 0.105 |
| SATHU | -0.020 | 0.053 | 0.713 | 0.913 | TC* | 0.006 | 0.061 | 0.127 | 0.266 |
| TC | -0.020 | 0.018 | 0.264 | 0.913 | TG | -0.046 | 0.077 | 0.549 | 0.610 |
| TG | -0.008 | 0.016 | 0.607 | 0.913 | VATSAT | 0.195 | 0.244 | 0.424 | 0.606 |
| VAT | 0.019 | 0.040 | 0.635 | 0.913 | VATSAT-adjBMI | 0.127 | 0.198 | 0.522 | 0.610 |
| VAT-adjBMI | 0.002 | 0.042 | 0.963 | 0.994 | WHR | -0.362 | 0.191 | 0.059 | 0.148 |
| VATHU | -0.012 | 0.048 | 0.803 | 0.913 | WHR-adjBMI | -0.287 | 0.140 | **0.041** | 0.148 |
| VATSAT | -0.017 | 0.044 | 0.707 | 0.913 | | | | | |
| VATSAT-adjBMI | -0.021 | 0.042 | 0.616 | 0.913 | | | | | |
| WHR | -0.008 | 0.013 | 0.546 | 0.913 | | | | | |
| WHR-adjBMI | -0.007 | 0.013 | 0.611 | 0.913 | | | | | |

P-values and q-values < 0.05 are in bold.

*MR results taken from Egger's method as (unbalanced) horizontal pleiotropy was significant.

^significant heterogeneity (p = 0.0217).

Table 5  Top 3 genes shared between SCZ/BD and CM traits

| | SCZ | | | | Bipolar disorder | | |
|---|---|---|---|---|---|---|---|
| **Trait** | **Top 3 genes** | **SNP** | **Tdr11** | **Trait** | **Top 3 genes** | **SNP** | **Tdr11** |
| **Adiponectin** | GNL3 | rs1108842 | 1.000 | **Adiponectin** | PBRM1 | rs2083180 | 0.925 |
| | ZNF664 | rs11057409 | 0.998 | | TWF2 | rs353547 | 0.863 |
| | SEMA3G | rs648514 | 0.993 | | ADIPOQ | rs12495941 | 0.805 |
| **BMI** | SLC39A8 | rs13107325 | 1.000 | **BMI** | LMX1B | rs7857133 | 0.975 |
| | NT5C2 | rs11191582 | 1.000 | | CADM2 | rs9829032 | 0.972 |
| | FOXO3 | rs9400239 | 1.000 | | GIPC2 | rs11162405 | 0.971 |
| **CAD** | PFN1P11 | rs11191416 | 1.000 | **CAD** | WBP1L | rs284863 | 0.769 |
| | FES | rs2521501 | 1.000 | | SERBP1P3 | rs73082037 | 0.724 |
| | NT5C2 | rs79780963 | 0.999 | | HDAC9 | rs12670036 | 0.715 |
| **DM** | MPHOSPH9 | rs1727313 | 0.990 | **DM** | PCBD2 | rs7728823 | 0.894 |
| | TLE1 | rs2796441 | 0.988 | | SFMBT1 | rs9865094 | 0.851 |
| | PROX1 | rs340835 | 0.986 | | ABCB9 | rs4275659 | 0.808 |
| **FG** | PROX1-AS1 | rs7529073 | 0.996 | **FG** | FEN1 | rs4246215 | 0.828 |
| | IFT172 | rs2272417 | 0.986 | | MRPL33 | rs4666024 | 0.691 |
| | DPYSL5 | rs10196501 | 0.975 | | BABAM2 | rs937813 | 0.619 |
| **FG adjBMI** | PROX1-AS1 | rs7529073 | 0.999 | **FG adjBMI** | FEN1 | rs4246215 | 0.754 |
| | SYNGAP1 | rs411136 | 0.988 | | MRPL33 | rs9678336 | 0.569 |
| | DPYSL5 | rs920435 | 0.983 | | BABAM2 | rs937813 | 0.568 |
| **Fat %** | TUFM | rs4788099 | 0.995 | **Fat %** | NEXN | rs1780050 | 0.983 |
| | CRTC1 | rs757318 | 0.995 | | SORT1 | rs1278664 | 0.978 |
| | DNAH10 | rs7133378 | 0.995 | | LINC01830 | rs2339519 | 0.918 |
| **HDL** | SLC39A8 | rs13107325 | 1.000 | **HDL** | PGAP3 | rs2517956 | 1.000 |
| | C12orf65 | rs1879379 | 1.000 | | TCAP | rs113612868 | 0.994 |
| | NFATC3 | rs8044995 | 1.000 | | VARS2 | rs4678 | 0.970 |
| **INS** | PSIP1 | rs10283923 | 0.889 | **INS** | ANKS1B | rs923724 | 0.780 |
| | ANKS1A | rs2820237 | 0.883 | | ANKS1A | rs847846 | 0.712 |
| | ANKS1B | rs923724 | 0.836 | | CEP68 | rs2723086 | 0.711 |
| **INS adjBMI** | GPN1 | rs10173720 | 0.946 | **INS adjBMI** | ANKS1A | rs847846 | 0.816 |
| | ANKS1A | rs2820237 | 0.941 | | TMEM131 | rs17424787 | 0.804 |
| | PPARG | rs6802898 | 0.940 | | TET2 | rs9884482 | 0.758 |
| **DM-2ndset** | KCNQ1 | rs234868 | 0.988 | **DM-2ndset** | OR12D3 | rs3749971 | 0.820 |
| | CDKAL1 | rs4712556 | 0.975 | | PPM1F | rs1005579 | 0.812 |
| | PSMD6 | rs3816157 | 0.945 | | ASAP2 | rs7605582 | 0.811 |
| **LDL** | STK19 | rs389883 | 1.000 | **LDL** | SP4 | rs12668354 | 0.956 |
| | SUGP1 | rs2074550 | 0.999 | | ATXN7L2 | rs2781553 | 0.945 |
| | SP4 | rs7811417 | 0.995 | | MED24 | rs12309 | 0.923 |
| **Leptin** | PTPN7 | rs9077 | 0.785 | **Leptin** | MLN | rs3828783 | 0.551 |
| | LINC02029 | rs900400 | 0.761 | | | | |
| | *MLN* | rs3828783 | 0.760 | | | | |
| **Leptin adjBMI** | *ADAMTSL3* | rs4374136 | 0.868 | **Leptin adjBMI** | *TRANK1* | rs4441609 | 0.620 |
| | *HSPA12A* | rs3010476 | 0.729 | | *TRPC4AP* | rs6120812 | 0.585 |
| | *YWHAE* | rs12936636 | 0.718 | | *PAX8* | rs2241975 | 0.532 |
| **TC** | *MSH5* | rs3117577 | 1.000 | **TC** | *PGAP3* | rs2517956 | 0.999 |
| | *TNXB* | rs1150753 | 1.000 | | *TCAP* | rs113612868 | 0.994 |
| | *PSORS1C1* | rs3130557 | 1.000 | | *SP4* | rs12668354 | 0.971 |
| **TG** | *MSH5-SAPCD1* | rs3131379 | 1.000 | **TG** | *MAP1LC3A* | rs6059908 | 0.980 |
| | *ATF6B* | rs1269852 | 1.000 | | *LINC00243* | rs886424 | 0.978 |
| | *PSORS1C1* | rs3130557 | 1.000 | | *GGT7* | rs11546155 | 0.952 |
| **WHR-** | *CCDC92* | rs4765219 | 0.998 | **WHR-** | *C1orf105* | rs4916261 | 0.870 |
| | *BCL11B* | rs1257432 | 0.994 | | *IKZF3* | rs907092 | 0.863 |
| | *WSCD2* | rs3764002 | 0.989 | | *HOTTIP* | rs17428025 | 0.855 |
| **WHR-adjBMI** | *ITIH1* | rs2710323 | 1.000 | **WHR-adjBMI** | *C1orf105* | rs6700880 | 0.934 |
| | *STAB1* | rs1010554 | 1.000 | | *GNL3* | rs1108842 | 0.920 |
| | *RFLNA* | rs11057409 | 0.999 | | *HLA-DMA* | rs9276931 | 0.878 |
| **DBP** | *SLC39A8* | rs13107325 | 1.000 | **DBP** | *CYP1A2* | rs11072506 | 0.961 |
| | *FURIN* | rs17514846 | 1.000 | | *ZSCAN12* | rs2232423 | 0.921 |
| | *BORCS7-ASMT* | rs11191454 | 0.999 | | *OR2B2* | rs61742093 | 0.915 |
| **SBP** | *CNNM2* | rs11191548 | 1.000 | **SBP** | *CYP1A2* | rs11072506 | 0.871 |
| | *CYP17A1* | rs17115100 | 1.000 | | *EDEM2* | rs3746429 | 0.723 |
| | *FURIN* | rs17514846 | 1.000 | | *NMT1* | rs4793172 | 0.664 |
| **PAT** | *CEBPA-AS1* | rs16967952 | 0.828 | **PAT** | *(no SNP with Tdr11 > 0.5)* | | |
| | *HS3ST4* | rs4787333 | 0.730 | | | | |
| | *PLAGL1* | rs9399468 | 0.700 | | | | |
| **PAT adjHtWt** | *CEBPA-AS1* | rs16967952 | 0.855 | **PAT adjHtWt** | *NEK4* | rs2072390 | 0.581 |
| | *OTX2-AS1* | rs762098 | 0.727 | | *LHB* | rs6509412 | 0.549 |
| | *NEK4* | rs2072390 | 0.712 | | | | |
| **SAT** | *PTPRK* | rs9375544 | 0.823 | **SAT** | *GSDMA* | rs3859192 | 0.835 |
| | *IZUMO1* | rs8106205 | 0.776 | | *PGAP3* | rs903501 | 0.637 |
| | *SPTSSB* | rs464766 | 0.747 | | *INO80D* | rs817998 | 0.554 |
| **SATHU** | *WDR82* | rs6445358 | 0.796 | **SATHU** | *MYRIP* | rs2679799 | 0.524 |
| | *SMG6* | rs2019872 | 0.785 | | | | |
| | *C16orf45* | rs222129 | 0.571 | | | | |
| **VAT** | *ARMC4* | rs11006760 | 0.786 | **VAT** | *(no SNP with Tdr11 > 0.5)* | | |
| | *CUL3* | rs11681451 | 0.723 | | | | |
| | *SERPINI1* | rs2055026 | 0.710 | | | | |
| **VAT adjBMI** | *EYS* | rs4710261 | 0.840 | **VAT adjBMI** | *(no SNP with Tdr11 > 0.5)* | | |
| | *PEPD* | rs3786897 | 0.805 | | | | |
| | *ALG12* | rs9616370 | 0.762 | | | | |
| **VATHU** | *WDR82* | rs6445358 | 0.878 | **VATHU** | *PHF7* | rs2272088 | 0.632 |
| | *PBRM1* | rs10510760 | 0.664 | | *ITIH3* | rs2240919 | 0.541 |
| | *SMG6* | rs2019872 | 0.628 | | | | |
| **VATSAT** | *EYS* | rs4710261 | 0.883 | **VATSAT** | *CADM2* | rs6781999 | 0.523 |
| | *ELP3* | rs3757894 | 0.852 | | | | |
| | *PEPD* | rs3786901 | 0.819 | | | | |
| **VATSAT adjBMI** | *ELP3* | rs3757894 | 0.823 | **VATSAT adjBMI** | *(no SNP with Tdr11 > 0.5)* | | |
| | *EYS* | rs4710261 | 0.817 | | | | |
| | *PEPD* | rs3786901 | 0.798 | | | | |

The above tables displays genes with mapping to SNPs that remained after LD-clumping. Mapping was performed by the program BioMart. Please refer to Table 1 for abbreviation of the metabolic traits. Please refer to Supplementary Tables for detailed results for all shared SNPs with $tdr_{11}>0.5$.

Table 6 Selected top pathways derived from SNPs shared between SCZ/BD and cardiometabolic traits

| Pathway | Freq | AvgRank | Present_In |
|---|---|---|---|
| **SCZ** | | | |
| Neuronal System | 11 | 40.5 | BMI , CAD , DM , HDL , DM-2nd_set , TC , TG , VAT , VAT-adjBMI , VATSAT-adjBMI , WHR |
| Insulin secretion - Homo sapiens (human) | 9 | 21 | DBP , DM , FG , FG-adjBMI , Fat-percentage , HDL , DM-2nd_set , SBP , TG |
| VEGF | 9 | 37.4 | BMI , CAD , LDL , PAT , TC , TG , VAT-adjBMI , WHR , WHR-adjBMI |
| Aldosterone synthesis and secretion - Homo sapiens (human) | 9 | 42 | BMI , CAD , FG , HDL , TC , TG , VAT , WHR , WHR-adjBMI |
| RXR and RAR heterodimerization with other nuclear receptor | 9 | 65.6 | BMI , CAD , DM , FG-adjBMI , HDL , LDL , TC , TG , WHR-adjBMI |
| Axon guidance | 8 | 34.8 | BMI , CAD , FG , Fat-percentage , Leptin , TG , VATSAT , VATSAT-adjBMI |
| Statin Pathway | 7 | 22.3 | BMI , CAD , Fat-percentage , HDL , LDL , TC , TG |
| Signaling by EGFR | 7 | 51.7 | BMI , FG , Fat-percentage , HDL , Leptin , WHR , WHR-adjBMI |
| Lipoprotein metabolism | 6 | 3.2 | CAD , Fat-percentage , HDL , LDL , TC , TG |
| Proton Pump Inhibitor Pathway, Pharmacodynamics | 6 | 29.8 | BMI , DBP , DM , FG , SAT , TG |
| EPO signaling | 6 | 39.2 | LDL , PAT , TC , TG , VAT-adjBMI , WHR-adjBMI |
| IL-7 signaling | 6 | 42.2 | LDL , PAT , TC , TG , VAT-adjBMI , WHR-adjBMI |
| Brain-Derived Neurotrophic Factor (BDNF) signaling pathway | 6 | 42.3 | BMI , CAD , Fat-percentage , HDL , Leptin , WHR |
| cAMP signaling pathway - Homo sapiens (human) | 6 | 48.5 | BMI , CAD , FG , HDL , Leptin , WHR |
| **Bipolar disorder** | | | |
| Chromatin modifying enzymes | 7 | 53.1 | BMI , HDL , DM-2ndset , LDL , PAT-adjHtWt , TC , TG |
| Alpha Linolenic Acid and Linoleic Acid Metabolism | 5 | 53.2 | FG , FG-adjBMI , HDL , LDL , TC |
| Antigen processing and presentation - Homo sapiens (human) | 4 | 12.2 | TC , TG , WHR , WHR-adjBMI |
| Cell adhesion molecules (CAMs) - Homo sapiens (human) | 4 | 17 | TC , TG , WHR , WHR-adjBMI |
| Amyloid fiber formation | 4 | 20.5 | HDL , DM-2ndset , TC , TG |
| HATs acetylate histones | 4 | 22.8 | HDL , DM-2ndset , TC , TG |
| ERCC6 (CSB) and EHMT2 (G9a) positively regulate rRNA expression | 4 | 23.5 | HDL , DM-2ndset , TC , TG |
| TGF-beta super family signaling pathway canonical | 4 | 27.5 | CAD , LDL , SAT , TC |
| Phagosome - Homo sapiens (human) | 4 | 28.5 | TC , TG , WHR , WHR-adjBMI |
| Positive epigenetic regulation of rRNA expression | 4 | 29.5 | HDL , DM-2ndset , TC , TG |

Freq: frequency of being top-listed across all CM traits; AvgRank, average rank of pathway if top-listed.